# Raman microscopy as a defect microprobe for hydrogen bonding characterization in materials used in fusion applications

Cédric Pardanaud[*,1], Younès Addab[1], Céline Martin[1], Pascale Roubin[1], Bernard Pegourié[2], Martin Oberkofler[3], Martin Köppen[4], Timo Dittmar[4], Christian Linsmeier[4].

[1] *Laboratoire PIIM, Aix-Marseille Université/CNRS, Avenue escadrille Normandie-Niemen, 13397, Marseille*

[2] *IRFM, C. E. A., 13108 Saint-Paul-lez-Durance, France*

[3] *Max-Planck-Institut für Plasmaphysik, Boltzmannstraße 2, D-85748 Garching, Germany*

[4] *Forschungszentrum Jülich GmbH, Institut für Energie- undKlimaforschung - Plasmaphysik, 52425 Jülich, Germany*

We present the Raman microscopy ability to detect and characterize the way hydrogen is bonded with elements that will be used for ITER's plasma facing components.

For this purpose we first use hydrogenated amorphous carbon samples, formed subsequently to plasma-wall interactions (hydrogen implantation, erosion, deposition...) occurring inside tokamaks, to demonstrate how this technique can be used to retrieve useful information. We pay attention in identifying which spectroscopic parameters are sensitive to the local structure ($sp^3/sp^2$) and which gives information on the hydrogen content using isothermal and linear temperature ramp studies on reference samples produced by plasma enhanced chemical vapor deposition. We then focus on the possibility to use this fast, non-destructive and non-contact technique to characterize the influence of hydrogen isotope implantation in few nanometers of graphite and beryllium as C is still used in the JT-60 tokamak and Be is used in JET and will be used as plasma-facing component in the future reactor ITER. We also pay attention on implantation in tungsten oxide which may be formed accidently in the machine.

# 1 Context of the study

Understanding and controlling hydrogen isotope retention in plasma-facing components of tokamaks (i.e. machines studying the possibility to make energy from nuclear fusion reactions by confining plasma magnetically), is a crucial safety issue because the international ITER reactor will use T+D, tritium being radioactive [1]. Being able to quantify how the plasma-facing components are processed subsequently to plasma-wall interactions (ionic implantation, erosion, deposition, diffusion, chemical bond formation, material migration...[1-4]) is then of importance.

To this aim, deuterium retention inside the Tore Supra tokamak (TS), able to perform long-time discharges, has been extensively studied during long time scales of operation (5 years) [5-11], the main plasma-facing component, called the toroidal pumped limiter, being made of carbon/carbon composite [12]. Post mortem analyses were used to quantify the amount of D trapped inside the materials. These analyses revealed that most of this D was trapped as deuterated amorphous carbon deposits (a-C:D) [13, 14], D being involved in chemical bonds such as $C(sp^3)$-H and $C(sp^2)$-H. However, the amount of trapped D was not the amount estimated by controlling how much D was introduced and how much was pumped out just after plasma discharges (in situ particle balance experiments). Ion beam analyses of many deposits revealed D depletion in the deepest zone of the carbon deposits (i.e. oldest zone of the carbon deposits). To understand this behavior, the importance of a long-term D release has been suggested and then characterized: D was found to be released slowly but significantly with time, this phenomenon strongly depending on the sample temperature [7]. This long term release was then able to reconcile post mortem and particle balance measurements [7, 9]. However carbon is no longer planned to be used for the plasma-facing components of ITER. Its plasma-facing components will be made of beryllium and tungsten. Some mixed materials that will modify the initial properties of the pristine materials will be formed due to the presence of:

- oxygen impurities,

- neon or nitrogen seeded in the machine, just before confinement is partially lost, to spray the heat loads on higher surfaces and prevent from damaging the plasma facing components,

- sputtered species,

- hydrogen isotopes…

As an example, tungsten oxide may be formed in some zones more or less efficiently, depending on the surface temperature. However, as carbon has been very well documented during the past years, it remains a reference system. In this work we give some examples on how to interpret Raman spectra in order to obtain information on how the hydrogen is bonded in materials. In a first part we compare the thermal stability of reference a-C:H layers to that of TS deposits by means of Raman spectroscopy recorded in situ. In a second part we compare the influence of hydrogen bombardment on pristine graphite, one kind of tungsten oxide and beryllium.

## 2 Raman microscopy of a-C:H

### 2.1 State of the art

The 1000 - 1800 cm-1 spectral window of carbonaceous materials containing $sp^2$ atoms is dominated by two bands: the G and D bands [15]. They contain information on disorder such as the size of aromatic domains, the density of defects… For amorphous carbons (a-C), they can give an indication on the $sp^2/sp^3$ ratio, and an estimation of the H content for hydrogenated carbons (a-C:H) [15-17]. Spectroscopic parameters generally used to probe the bonding structure are: the width and the position of the G band, denoted $\Gamma_G$ and $\sigma_G$ respectively, the height ratio of the D and G bands, $H_D/H_G$, and the $m/H_G$ parameter with m being the slope of a superimposed photoluminescence background measured between 800 and 2000 cm$^{-1}$. The way they are determined on raw spectra is shown in the figure 1.

### 2.2 Post-mortem analyses of TS deposits

Figure 2 displays the $\Gamma_G$ parameter as a function of the $\sigma_G$ parameter for samples collected on the toroidal pumped limiter and on the neutralizer element of Tore Supra [14]. They display a continuous curve, ranging from nano-crystalline graphite ($\sigma_G \approx 1590$ cm$^{-1}$, $\Gamma_G \approx 25$-$50$ cm$^{-1}$) to amorphous carbon ($\sigma_G \approx 1510$ cm$^{-1}$, $\Gamma_G \approx 200$ cm$^{-1}$) with various changes in local order in between. Samples from the neutralizer range from nano-crystalline to amorphous whereas samples from the limiter are amorphous. However the former are the most ordered. This can be understood because the limiter is actively cooled during discharges, limiting the surface temperature variations subsequently to heat loads, whereas the neutralizer's deposits are not cooled due to loosely attached and thick deposits, allowing them to reach higher temperatures and then to get a higher degree of organization and a larger H release. As these measurements are rapid compared to ion beam analyses or other relevant techniques, it is possible to perform a statistical treatment on many samples coming from many places of the plasma-facing components. The conclusion is that deuterium retained in deposits found on the neutralizer can be neglected compared to that found on the deposition zones of the limiter. Note that the TS samples' spectroscopic parameters have been compared to those of a-C:H, ta-C:H and ta-C found in literature ([18] and references therein).

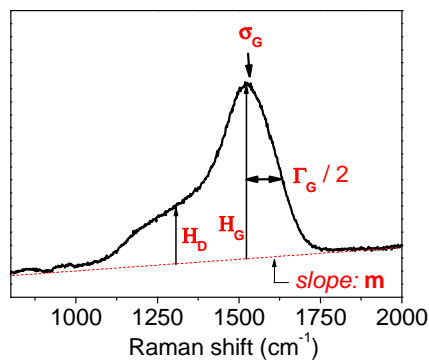

*Figure 1. Raman spectrum of a typical plasma-deposited amor-phous carbon. $\Gamma_G$ and $\sigma_G$ are the width at half maximum and the position of the G band, respectively. $H_D/H_G$ is the height ratio of the D and G bands. m is the slope of the photoluminescence background.*

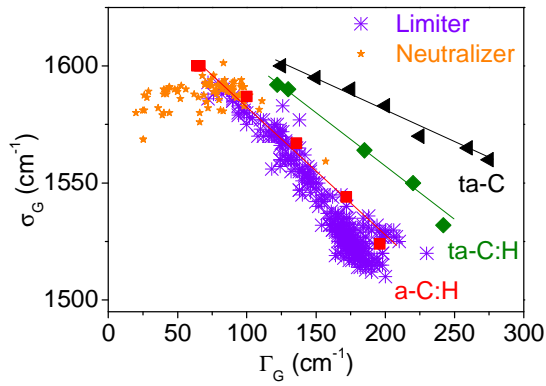

*Figure 2. G band wavenumber as a function of the G band width for different carbon samples. a-C:H, ta-C:H and ta-C are plasma deposited samples (the symbol t means tetrahedral, see reference [18] and references therein) whereas the samples referred as limiter and neutralizer refer to Tore Supra deposits.*

## 2.3 Thermal stability of reference and TS samples

Figure 3 displays the Raman spectra of one a-C:H sample that has been heated from room temperature to 1030°C ($H_G$ is normalized to 1 after a linear background subtraction). Temperature increase leads to local organization as evidenced by the G band narrowing and blue shift, and the increase of the D band. Temperature increase also leads to hydrogen release: the inset gives the linear relation we found previously between $H_D/H_G$ and the H content measured by ion beam analysis [19]. This linear relation was found with all kinds of hydrogen bonding (i.e. H bonded to $C(sp^2)$ or to $C(sp^3)$ atoms) and can thus be used as a tool to determine the H content.

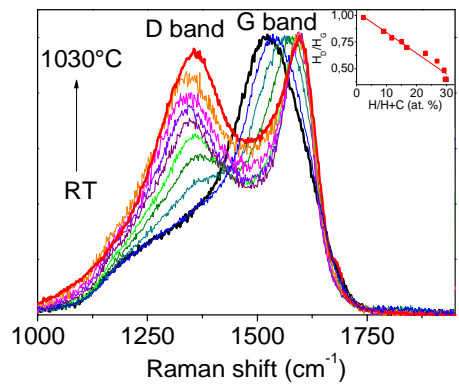

*Figure 3. Raman spectra of one a-C:H layers heated from room temperature to 1030°C (H/H+C=30% for the as deposited sample). The inset represents the linear relation found between $H_D/H_G$ and the H content obtained when heating the as deposited sample [19].*

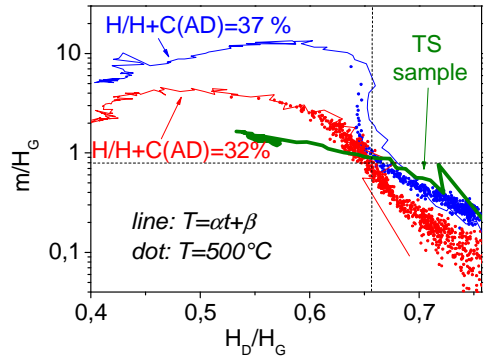

**Figure 4**. $m/H_G$ evolution under 500°C isotherm and linear ramp heating plotted against $H_D/H_G$ for two reference a-C:H samples and compared to TS deposits. AD means As Deposited.

Figure 4 displays the evolution of $m/H_G$ and $H_D/H_G$ for reference samples compared to TS samples under heating (both linear temperature ramp with $\alpha=3K.min^{-1}$ and isothermal heating at 500°C [20]). Samples were heated under argon atmosphere, spectra being recorded in situ in real time through a window. We see that data corresponding to the linear ramp and isotherms are superimposed in a quite good agreement. $H_D/H_G$ probes the hydrogen content whereas $m/H_G$ probes successively with increasing temperature: the $sp^2$ clustering/organization and then the H content of H bonded to $C(sp^3)$. Consequently, the data points can be separated in two zones in the $m/H_G$ versus $H_D/H_G$ plot: the upper left corner corresponding mainly to $sp^2$ clustering and the lower right corner corresponding to the release of H bonded to $C(sp^3)$. Note that the curves of the two reference samples have the same shapes in a plot where the x-axis is temperature, but are shifted in temperature which allows them to be clearly distinguished (not shown on this figure). The reference sample data presented in figure 4 were recorded in the thermal range where H comes mainly from the scission of $H-C(sp^3)$ bonds. Then, one can see that in the lower right corner of this figure the relation between $log10(m/H_G)$ and $H_D/H_G$ is linear ; both parameters give the same information, however they do not superimpose as the amount of hydrogen bonded to $sp^2$ carbons may not be the same. The curve corresponding to the TS sample behaves like that of the as deposited (AD) H/H+C=32% sample concerning the $sp^2$ clustering, whereas it behaves like the AD H/H+C=37% sample concerning the thermal stability.

## 3 Raman microscopy of Be:D and $WO_{3-x}$:D

Due to tritium retention, carbon has not been chosen for the design of the ITER wall components. They will be composed of Be and W. First results on the JET ITER like wall have revealed that tritium will be stored mainly in beryllium deposits [21].

Figure 5 displays a comparison of Raman spectra cor-responding to various pristine and D implanted samples, to check the sensitivity of the technique to ion implantation. The samples are Highly Oriented Pyrolitic Graphite (HOPG), tungsten oxide and polycrystalline beryllium. The energies and fluencies of the bombarding D ion were respectively in the range of ≈ 0.5-2 keV and ≈ $10^{17}$-$10^{19}$ ions/cm$^2$ ([22] for HOPG and W), depending on the material, allowing ions to be implanted up to a few nm to ≈ 10 nm in depth and creating a sufficiently high amount of defects so that the corresponding Raman signature is intense enough to be detected. One can see that the HOPG becomes amorphous due to implantation, as the spectrum, which has been modified subsequently to ion implantation, looks like that of figure 1, being composed of broad D and G bands. A narrower D band also appears at 1360 cm$^{-1}$, related to another kind of defect. The underlying unmodified HOPG signature is also seen [23]. The tungsten oxide sample has been prepared by heating a polycrystalline W sample at 600°C during 60 minutes under 1.5 bar of $O_2$. D implantation broadened the stretching (≈ 200-500 cm$^{-1}$) and bending (≈ 600-800 cm$^{-1}$) modes, changed their relative intensity ratio and two new bands appeared at ≈ 400 and 960 cm$^{-1}$. The band at 960 cm$^{-1}$ is known to be a surface W=O bond, suggesting that porosities may have been formed during bombardment.

The Be stretching mode, giving rise to a Raman active mode at 458 cm$^{-1}$, is also modified by bombardment. This band is shifted and broadened which is the sign of stress caused by implantation.

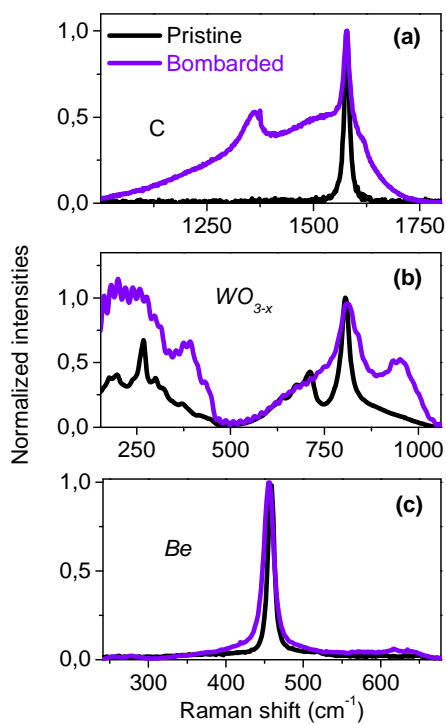

***Figure 5.*** *Normalized Raman spectra of pristine and D implanted graphite (a), $WO_{3-x}$ (b) and Be (c).*

# 3 Conclusions

Understanding and controlling hydrogen isotope retention in plasma-facing components of tokamaks is important due to safety issue. In this work, we have studied the Raman microscopy ability to detect and characterize the way hydrogen is bonded with elements that have been used for nowadays tokamaks' plasma-facing components (carbon) and that will be used for ITER's plasma-facing components (beryllium and tungsten).

Using continuous changes with temperature of various carbonaceous samples, we have shown how Raman spectroscopic parameters can be used to determine structural and H-content evolution. We have then compared the thermal stability of reference a-C:H deposits to deposits collected in the Tore Supra tokamak to investigate deeper how hydrogen is bonded.

We have evidenced that Raman spectroscopy is a sensitive non contact technique which can give information on how pristine tokamak material have been modified due to plasma/wall interactions. We have then evidenced that implantation modifies the vibrational properties of various materials that may be found in ITER (tungsten oxide and beryllium). We point out that the depth where structural or chemical modification occurs in the material, from a few nm to ≈ 10 nm, can be easily probed by means or Raman spectroscopy. Further studies are now needed to allow Raman spectroscopy to become a H-content quantitative method for tungsten oxide and beryllium.

*Acknowledgements.* *Part of this work has been carried out within the framework of the EUROfusion Consortium and has received funding from the European Union's Horizon 2020 research and innovation programme under grant agreement number 633053. The views and opinions expressed herein do not necessarily reflect those of the European Commission. We also ac-knowledge the EURATOM program, and the Federation de Recherche FR-FCM. C. P. wants to thank W. Jacob, T. Schwarz-Selinger, C. Hopf and G. Cartry.*